\documentclass[aps,prb,twocolumn,superscriptaddress]{revtex4-1}

\usepackage[pdftex,colorlinks,bookmarks = true]{hyperref} 
\usepackage[capitalize]{cleveref} 

\usepackage[usenames,dvipsnames]{xcolor}
\usepackage[caption=false]{subfig} 

\usepackage{csquotes} 
\usepackage{upgreek} 
\usepackage[version=3]{mhchem} 

\usepackage{amsmath}
\usepackage{amsfonts}
\usepackage{amssymb}
\usepackage{empheq} 
\usepackage{mathtools} 

\usepackage{siunitx} 
\usepackage{braket} 

\newcommand{\avg}[1]{\left\langle #1 \right\rangle} 
\let\vaccent=\v 
\renewcommand{\v}[1]{\boldsymbol{\mathbf{#1}}} 
\newcommand{\e}[1]{\, \mathrm{e}^{#1}} 
\renewcommand{\i}{\mathrm{i}} 
\newcommand{\cpi}{\uppi} 
\providecommand*{\groupU}[1]{\mathrm{U}(#1)} 
\providecommand*{\groupZ}[1]{\mathbb{Z}_{#1}} 
\providecommand*{\groupUZ}{\groupU{1} \times \groupZ{2}} 

\makeatletter 
\providecommand*{\diff}
	  {\@ifnextchar^{\DIfF}{\DIfF^{}}}
  \def\DIfF^#1{%
	  \mathop{\mathrm{\mathstrut d}}%
		  \nolimits^{#1}\gobblespace}
  \def\gobblespace{%
	  \futurelet\diffarg\opspace}
  \def\opspace{%
	  \let\DiffSpace\!%
	  \ifx\diffarg(%
		  \let\DiffSpace\relax
	  \else
		  \ifx\diffarg[%
			  \let\DiffSpace\relax
		  \else
			  \ifx\diffarg\{%
				  \let\DiffSpace\relax
			  \fi\fi\fi\DiffSpace}
\makeatother

\begin{document}

\title{Time reversal symmetry breakdown in normal and superconducting states in frustrated three-band  systems}

\author{Troels Arnfred Bojesen}
\affiliation{Department of Physics, Norwegian University of Science and Technology, NO-7491 Trondheim, Norway}
\author{Egor Babaev}
\affiliation{Department of Theoretical Physics, The Royal Institute of Technology, 10691 Stockholm, Sweden}
\affiliation{Physics Department, University of Massachusetts, Amherst, Massachusetts 01003, USA}
\author{Asle Sudb\o{}}
\affiliation{Department of Physics, Norwegian University of Science and Technology, NO-7491 Trondheim, Norway}

\date{\today}

\begin{abstract}
We discuss the phase diagram and phase transitions in $U(1)\times \groupZ{2}$ three-band superconductors with broken time reversal symmetry. 
We find that  beyond mean field approximation and for sufficiently strong frustration of interband interactions there appears an unusual 
metallic state precursory to a superconducting phase transition. In that state, the system is not superconducting. Nonetheless, it  features 
a  spontaneously broken $\groupZ{2}$ time reversal symmetry. By contrast, for weak frustration of interband coupling the energy of a domain  
wall between different  $\groupZ{2}$ states is low and  thus fluctuations restore broken time reversal symmetry in the superconducting state 
at low temperatures.
\end{abstract}

\maketitle


In recent years, the discovery of superconductors such as the Iron Pnictides \cite{iron}, has generated much interest for multiband 
superconducting systems. From a theoretical viewpoint, one of the main reasons for the strong interest is that in contrast to previously known
two-band materials, iron-based superconductors  may exhibit dramatically different physics due to the possibility of {\it frustrated} 
inter-band Josephson couplings originating with more than two bands crossing the Fermi-surface \cite{nagaosa,zlatko,johan3,maiti}. In two-band 
superconductors the Josephson coupling locks the phase differences between the bands to 0 or $\cpi$. By contrast, if one has three bands and 
the frustration of interband coupling is sufficiently strong, the ground state interband phase difference can be  different from $0$ or $\cpi$. 
This leads to a  superconducting state which breaks time reversal symmetry (BTRS) \cite{nagaosa,zlatko}. From a symmetry viewpoint such a ground 
state breaks $\groupUZ$ \cite{johan3}. Recently, such a scenario has received solid theoretical support \cite{maiti} in connection with 
hole-doped \ce{Ba_{$1-x$}K_{$x$}Fe2As2}. The possibility of this new physics arising also in  other classes of materials  is currently under 
investigation \cite{agterberg2011}. For other scenarios of time reversal symmetry breakdown in iron-based superconductors discussed in the 
literature, see \cite{other1,other2}.

Three band superconductors with frustrated interband Josephson couplings feature several properties that are radically different from their two-band 
counterparts. These include (I) the appearance of a massless so-called Leggett mode at the $\groupZ{2}$ phase transition \cite{lin}; 
(II) the existence of new mixed phase-density collective modes in the state with broken time-reversal symmetry (BTRS) \cite{johan3,stanev,maiti} 
in contrast to  the \enquote{phase-only} Leggett  collective mode in two-band materials \cite{leggett}; and (III) the existence of 
(meta-)stable excitations characterized by $\mathbb{CP}^2$ topological invariants \cite{cp21,cp22}.

So far the phase diagram of frustrated  three-band  superconductors has been investigated only at the mean-field level \cite{zlatko,maiti}. However, 
the iron-based materials feature relatively high $T_c$, as well as being far from the type-I regime.  Hence, one may expect fluctuations to be of 
importance. 

In this paper, we study the phase diagram  of a three-band superconductor in two spatial dimensions in the London limit, beyond
mean-field approximation. The results should apply to relatively thin films of iron-based superconductors where, owing to low dimensionality, 
fluctuation effects are particularly important. The main findings of this work are as follows. (I) When the  frustration is sufficiently 
strong, the phase diagram acquires an unusual fluctuation-induced metallic state which is a precursor to the BTRS superconducting phase. 
This metallic state exhibits a broken $\groupZ{2}$ time-reversal symmetry. A salient feature is that, although the state is metallic and 
non-superconducting, it nevertheless features a persistent interband Josephson current in momentum space which breaks time reversal 
symmetry. (II) When the frustration is weak (i.e. when phase differences are only slightly different from 0 or $\pi$) we find that the 
system can undergo a fluctuation driven restoration of the $\groupZ{2}$ symmetry at very low temperatures. 

The London model for a three-band superconductor is given by 
\begin{widetext}
  \begin{equation}
F= 
\sum_{\alpha=1,2,3}\frac{|\psi_\alpha|^2}{2}(\nabla \theta_\alpha -e \v{A})^2 
+ \sum_{\alpha,\alpha'>\alpha } \eta_{\alpha\alpha'}|\psi_\alpha||\psi_{\alpha'}|\cos(\theta_\alpha-\theta_{\alpha'} )
+\frac{1}{2}(\nabla \times \v A)^2.
\label{london}
\end{equation}
\end{widetext}
Here,  $|\psi_\alpha|\e{\i \theta_\alpha}$ denotes the superconducting condensate components in different bands labeled by $\alpha=1,2,3$, 
while the second term represents interband Josephson couplings.  The field $\v A$ is the magnetic vector potential that couples minimally 
to the charged condensate matter fields. By collecting gradient terms for phase differences, it can also be rewritten as 
\begin{widetext}
\begin{equation}   
F= \frac{1}{2\varrho^2}\left(\sum_\alpha |\psi_\alpha|^2\nabla \theta_\alpha -e \varrho^2{\bf A}\right)^2  +  \frac{1}{2}(\nabla \times {\bf A})^2+ 
\sum_{\alpha,\alpha'>\alpha}\frac{|\psi_\alpha|^2|\psi_{\alpha'}|^2}{2\varrho^2}[\nabla(\theta_\alpha-\theta_{\alpha'} )]^2
 +\eta_{\alpha\alpha'}|\psi_\alpha||\psi_{\alpha'}|\cos(\theta_\alpha-\theta_{\alpha'} ),
\label{ps}		
\end{equation}
\end{widetext}
where $\varrho^2=\sum_\alpha |\psi_\alpha|^2$. {This shows that the vector potential is coupled only to the $U(1)$ sector of the model, and 
not to phase differences.}

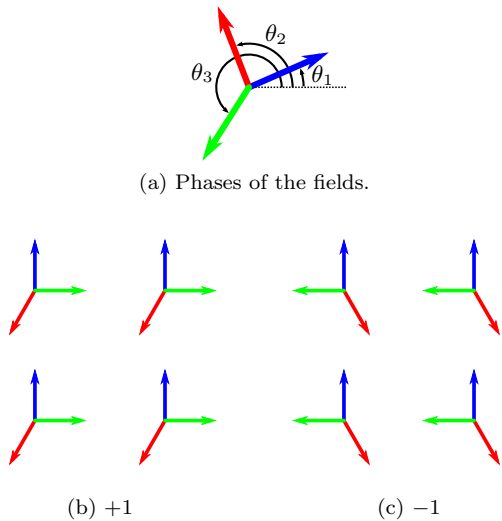
\begin{figure}[ht]
  \subfloat[Phases of the fields.\label{fig:phase_definition}]{
    \def\svgwidth{0.4\columnwidth}
      \begingroup%
    \makeatletter%
    \providecommand\color[2][]{%
      \errmessage{(Inkscape) Color is used for the text in Inkscape, but the package 'color.sty' is not loaded}%
      \renewcommand\color[2][]{}%
    }%
    \providecommand\transparent[1]{%
      \errmessage{(Inkscape) Transparency is used (non-zero) for the text in Inkscape, but the package 'transparent.sty' is not loaded}%
      \renewcommand\transparent[1]{}%
    }%
    \providecommand\rotatebox[2]{#2}%
    \ifx\svgwidth\undefined%
      \setlength{\unitlength}{135.16518555bp}%
      \ifx\svgscale\undefined%
	\relax%
      \else%
	\setlength{\unitlength}{\unitlength * \real{\svgscale}}%
      \fi%
    \else%
      \setlength{\unitlength}{\svgwidth}%
    \fi%
    \global\let\svgwidth\undefined%
    \global\let\svgscale\undefined%
    \makeatother%
    \begin{picture}(1,0.59557126)%
      \put(0,0){\includegraphics[width=\unitlength]{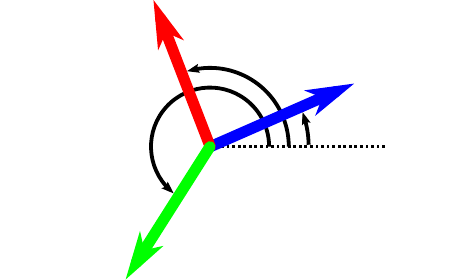}}%
      \put(0.68813623,0.30221329){\color[rgb]{0,0,0}\makebox(0,0)[lb]{\smash{$\theta_{1}$}}}%
      \put(0.5098767,0.4595013){\color[rgb]{0,0,0}\makebox(0,0)[lb]{\smash{$\theta_{2}$}}}%
      \put(0.31188689,0.31269923){\color[rgb]{0,0,0}\makebox(0,0)[rb]{\smash{$\theta_{3}$}}}%
    \end{picture}%
  \endgroup%
  }\\
  \subfloat[$+1$]{\includegraphics[width=0.4\columnwidth]{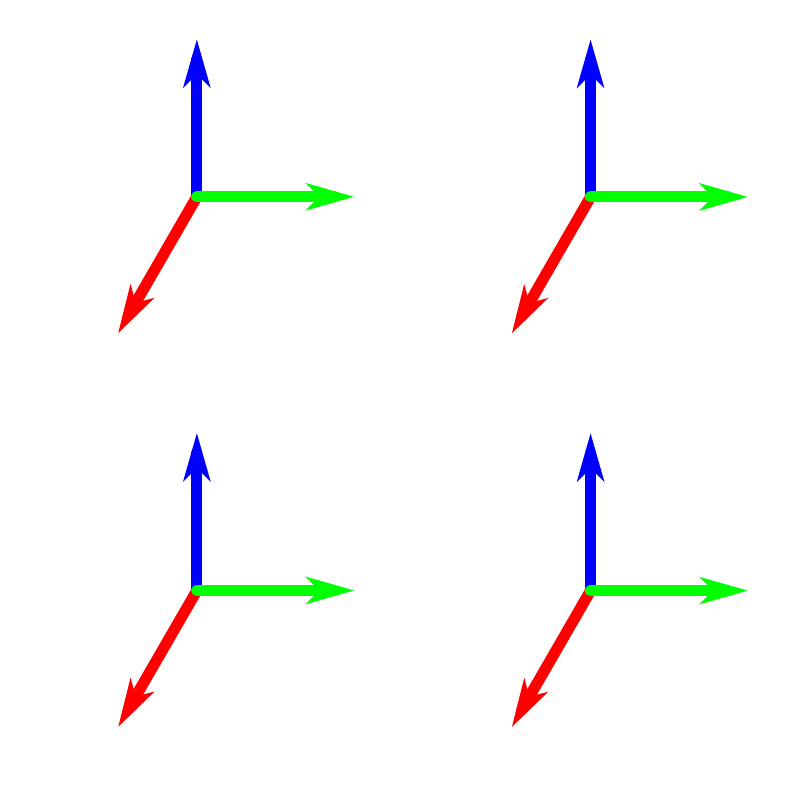}}
  \qquad
  \subfloat[$-1$]{\includegraphics[width=0.4\columnwidth]{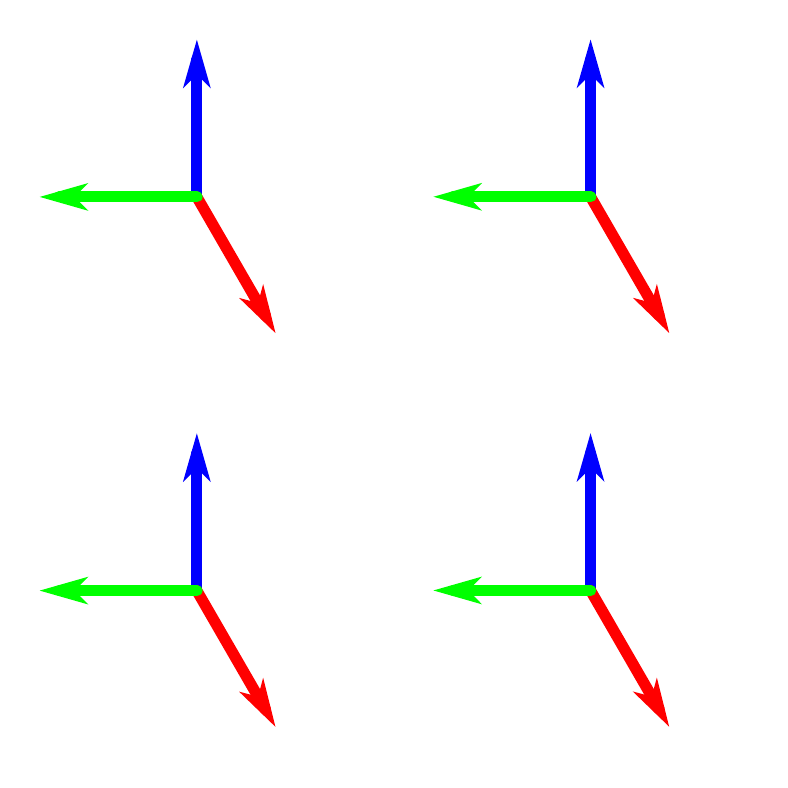}}
  \caption{(Colors online) (b) and (c) are examples of phase configurations for the two $\groupZ{2}$ symmetry classes of the ground states, shown 
at a $2\times 2$ lattice of selected points of the $xy$ plane. Here $g_{12} > g_{23} > g_{13} > 0$. The spatial contribution to the energy is 
minimized by making the spatial gradient zero (hence breaking the global $\groupU{1}$ symmetry). Then there are two classes of phase configurations, 
one with chirality +1 and one with chirality -1, minimizing the energy associated with the interband interaction. The chirality is defined as $+1$ 
if the phases (modulo $2\cpi$) are cyclically ordered $\theta_1<\theta_2<\theta_3$, and $-1$ if not. The arrows $(\textcolor{blue}{\longrightarrow},\textcolor{red}{\longrightarrow},\textcolor{green}{\longrightarrow})$ correspond to 
$(\theta_1,\theta_2,\theta_3)$, as shown in (a). }
  \label{fig:gr_state_Z2}
\end{figure}

When the Josephson couplings $\eta_{\alpha\alpha'}$ are positive, each Josephson term by itself prefers to lock phase difference 
to $\cpi$, i.e. $\theta_\alpha-\theta_{\alpha'} =\cpi $. Since this is not possible for three phases, the system is frustrated. The 
system breaks time reversal symmetry when Josephson couplings are minimized by two inequivalent phase lockings, shown in 
Fig. \ref{fig:gr_state_Z2}. The phase lockings are related by complex conjugation of the fields $\psi_\alpha$. Thus, by choosing 
one of these phase locking patterns the system breaks time reversal symmetry \cite{nagaosa,zlatko,johan3}. 

\begin{figure}[ht] 
  \subfloat[A $\groupZ{2}$ broken, $\groupU{1}$ symmetric configuration with $+1$ chirality.]{\includegraphics[width=0.4\columnwidth]{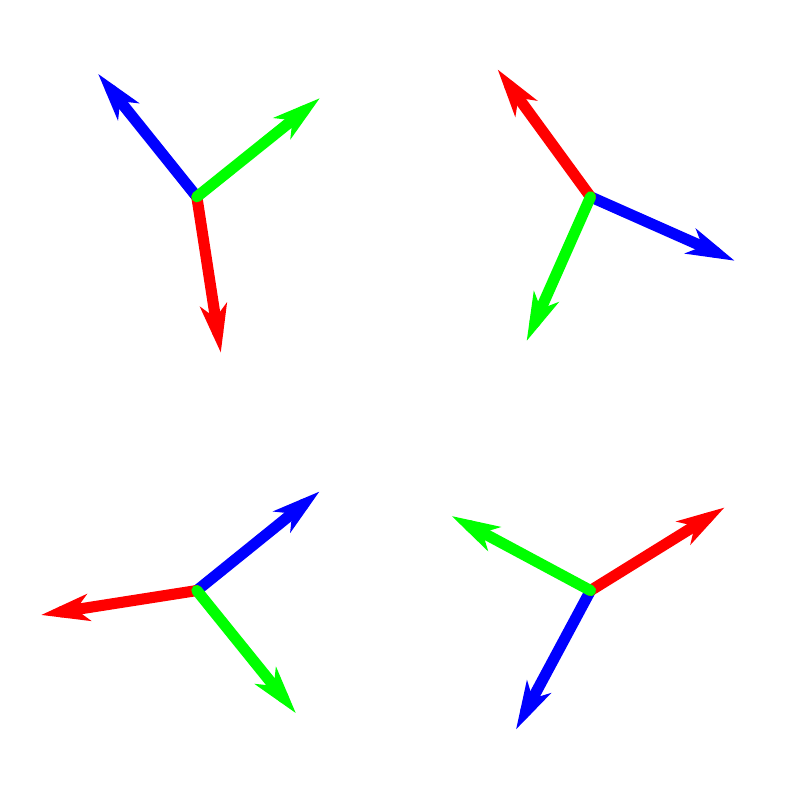}}
  \qquad
  \subfloat[A $\groupZ{2}$ and $\groupU{1}$ symmetric configuration.]{\includegraphics[width=0.4\columnwidth]{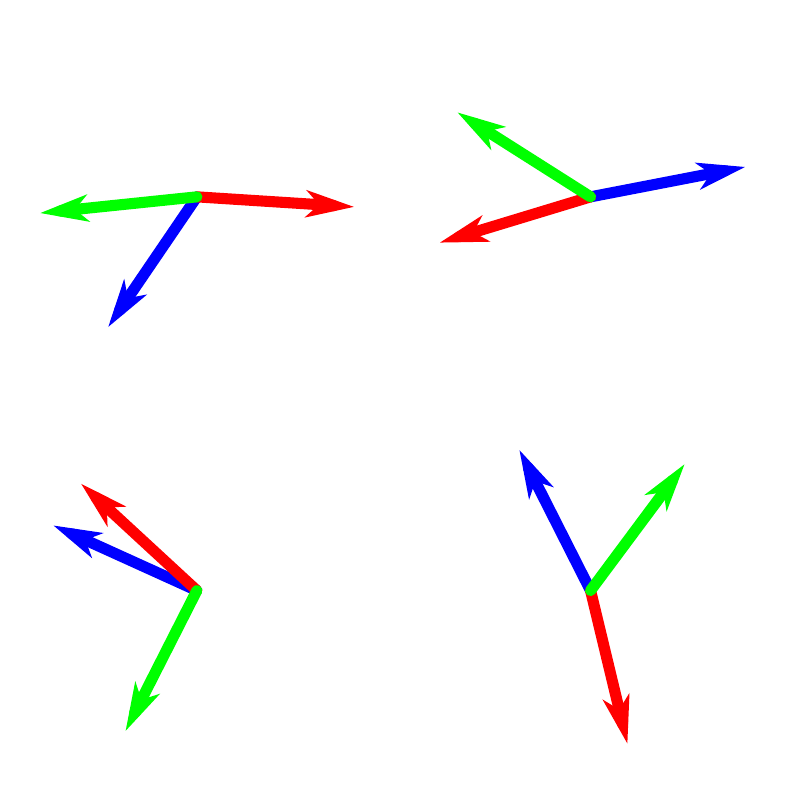}}
  \caption{(Colors online) A schematic illustration of phase configurations in the normal state which break time reversal symmetry vs the normal state which does not. 
In (a) the $\groupZ{2}$ symmetry is broken (and of +1 chirality) since the phases of all lattice sites have the same cyclic ordering. There is however no spatial 
ordering of the phases, hence the configuration is $\groupU{1}$ symmetric. In (b) neither the $\groupZ{2}$ nor the $\groupU{1}$ symmetry is broken. 
The arrows $(\textcolor{blue}{\longrightarrow},\textcolor{red}{\longrightarrow},\textcolor{green}{\longrightarrow})$ correspond to $(\theta_1,\theta_2,\theta_3)$, 
as shown in \cref{fig:phase_definition}.}
  \label{fig:configurations}
\end{figure}

In this work, we address the phase transitions in a two dimensional three-band superconductor with broken time-reversal symmetry. 
A Berezinskii-Kosterlitz-Thouless (BKT) phase transition in $\groupU{1}$ systems is driven by proliferation of vortex-antivortex pairs, 
while an Ising phase transition is driven by proliferation of $\groupZ{2}$ domain walls. The nontriviality of the problem of phase transitions 
in the three-band model is due to the spectrum of topological excitations of the model. Firstly, the model features singly-quantized 
composite vortices where all phases wind by $2\pi$, i.e. $\Delta \theta_1  \equiv \oint \nabla \theta_1=2\pi, \Delta \theta_2=2\pi,\Delta \theta_3=2\pi$. 
We will denote them (1,1,1). As is clear from Eq. \eqref{ps}, such a vortex has topological charge only in the $\groupU{1}$ sector of the model. 
It has no phase winding in the phase differences and thus does not carry topological charge in $\groupZ{2}$ sector. In addition, the model features 
other topological defects discussed in detail in  Refs. \onlinecite{cp21,cp22}. These are $\groupZ{2}$ domain walls (several solutions with different 
energies), fractional-flux vortices with linearly divergent energy, as well as $\mathbb{CP}^2$ skyrmions which are combined vortex-domain wall defects carrying 
topological charges in both the $\groupU{1}$ and $\groupZ{2}$ sectors of the model. This spectrum of topological excitations distinguishes this model 
from other $\groupUZ$ systems, like e.g. $XY$-Ising model \cite{PhysRevLett.66.1090}. The model is also principally different from  $[\groupU{1}]^3$ 
superconductors, since in such systems fractional vortices have logarithmically divergent energy and thus drive BKT phase transitions 
\cite{smiseth,npb}. The difference between a  $\groupZ{2}$-ordered and disordered state is illustrated in Fig. \ref{fig:configurations}.

In two dimensions, the effective magnetic field penetration length is inversely proportional to the film thickness \cite{pearl}. We thus begin by 
discussing the limit of very large penetration length, in which we may neglect the coupling to the vector potential. We discuss the phase diagram 
of the model in the case of a finite penetration length in our summary. 

The partition function of the lattice version of the model \eqref{london} reads
 \begin{equation}
 Z =  \prod_{\alpha, i} \left[\int_{-\cpi}^{\cpi} \frac{\diff \theta_{\alpha,i}}{2\cpi}\right]  \exp(-\beta H),
 \end{equation}
 where the Hamiltonian is given by
 \begin{multline}
 H = -\sum_{\avg{i,j},\alpha} \cos\left(\theta_{\alpha,i} - \theta_{\alpha,j}\right) \\
 + \sum_{i,\alpha' > \alpha} g_{\alpha \alpha'} \cos\left(\theta_{\alpha,i} - \theta_{\alpha',i} \right).
 \label{eq:fundamental_action}
 \end{multline}
 $i,j \in \set{1,2,\ldots,N=L \times L}$ denote sites on a lattice of size $L \times L$ and $\avg{i,j}$ indicates nearest neighbor lattice sites 
(assuming periodic boundary conditions). $\beta$ is the (properly rescaled) coupling (\enquote{inverse temperature}) and 
$g_{\alpha \alpha'}$ are interband Josephson couplings. Here, we consider the case of similar prefactors for the three gradient terms.

{Algebraically decaying correlations and frustration effects typically render two dimensional $\groupUZ$-symmetric models difficult to investigate  numerically \cite{Hasenbusch_et_al}.} In this work we emply a non-equilibrium approach, namely that of short time critical dynamics (STCD). See e.g. the review articles \onlinecite{Ozeki_Ito_2007} and 
\onlinecite{Albato_et_al_2011}, and references therein. See online supplementary material for details\footnote{In the results presented in this paper, we have mostly used an order parameter measuring phase-ordering in individual phases $\theta_1,\theta_2,\theta_3$. Other choices do however exist in this multi-component case, such as measuring order in $\theta_1+ \theta_2$, $\theta_1+ \theta_3$, and $\theta_2+ \theta_3$. We may also use an order parameter which is an average over all of the above mentioned possible choices. We have checked all these cases, and find no difference in the phase diagram.}

\begin{figure}[ht]
\includegraphics{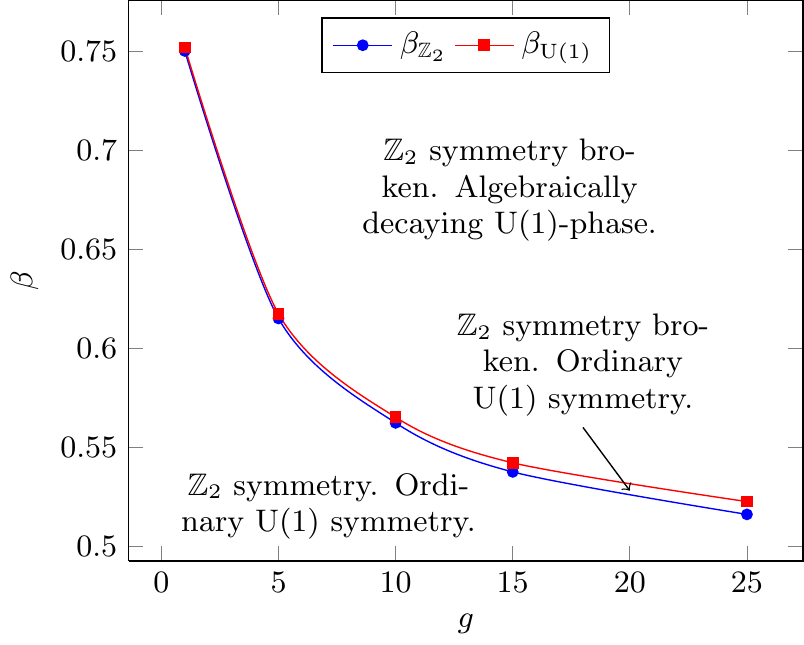}
\caption{(Color online) Phase diagram for the three-{band} model with $g_{12} = g_{23} = g_{31} = g$. $g \in \left[1,\ldots,25\right]$. The $\beta_{\groupU{1}}$ 
line lies above the $\beta_{\groupZ{2}}$ line for the investigated values of $g$. Error bars are smaller than symbol sizes. 
Lines are guides to the eye.}
\label{fig:pd_gen_max_sym}
\end{figure}

First, we consider the case $g_{12} = g_{23} = g_{31} = g$, which is shown in Fig. \ref{fig:pd_gen_max_sym}. The phase transitions for the 
$\groupZ{2}$ and $\groupU{1}$ symmetries are close, but clearly separated for all values of $g$. This means that beyond the mean-field approximation 
there appears a new phase. As the temperature increases from the low-temperature maximally ordered phase, an unbinding of  vortex-antivortex pairs 
of composite vortices first takes place.  In the resulting state the free composite vortices $(1,1,1)$ and $(-1,-1,-1)$ do not further decompose  
into $(1,0,0)$, $(0,1,0)$, $(0,0,1)$ fractional vortices, because Josephson coupling provides linear confinement of the constituent fractional 
vortices \cite{cp22}). Due to this confinement, the proliferation of $(1,1,1)$ and $(-1,-1,-1)$ vortex-antivortex pairs disorders only the 
$\groupU{1}$ sector of the model described by the first term in  Eq. \eqref{ps}. However, these defects  do not restore $\groupZ{2}$ symmetry. 
{This results in a formation of a new state which  is non-superconducting 
but breaks  broken time-reversal symmetry. The effective
model which describes this new state is given by the last terms in Eq. \eqref{ps},}
\begin{multline}
 F_{\groupZ{2}} = \sum_{\alpha,\alpha'>\alpha}\Bigl\{ \frac{|\psi_\alpha|^2|\psi_{\alpha'}|^2}{2\varrho^2}[\nabla(\theta_\alpha-\theta_{\alpha'} )]^2 \\
  +\eta_{\alpha\alpha'}|\psi_\alpha||\psi_{\alpha'}|\cos(\theta_\alpha-\theta_{\alpha'} ) \Bigr\}
  \label{z2}
\end{multline}

Secondly, at higher temperatures the $\groupZ{2}$ domain walls proliferate and restore the symmetry completely. The physical interpretation of this 
precursor normal state with broken time reversal symmetry is as follows. In the BTRS superconducting state there is a ground state phase difference other 
than 0 or $\cpi$ between components. This implies the existence of persistent interband Josephson currents. Two different $\groupZ{2} $ phase locking patterns 
mean that there are two inequivalent interband Josephson current \enquote{loops in $\bf k$-space}. Namely, one loop is of the type band 1 $\to$ band 2 $\to$ 
band 3 $\to$ band 1, the other is of the type band 1 $\to$ band 3 $\to$ band 2 $\to$ band 1. The non-superconducting $\groupZ{2}$-ordered phase corresponds 
to the situation where superconducting phases exhibit exponentially decaying correlations due to proliferation of vortex-antivortex pairs. 
{What sets this  state  apart from the situation found in conventional superconductors is that the three-band system
retains a persistent interband Josephson current in $\v k$-space which breaks the time reversal symmetry}, see Fig. \ref{fig:configurations}.

\begin{figure}[ht]
\includegraphics{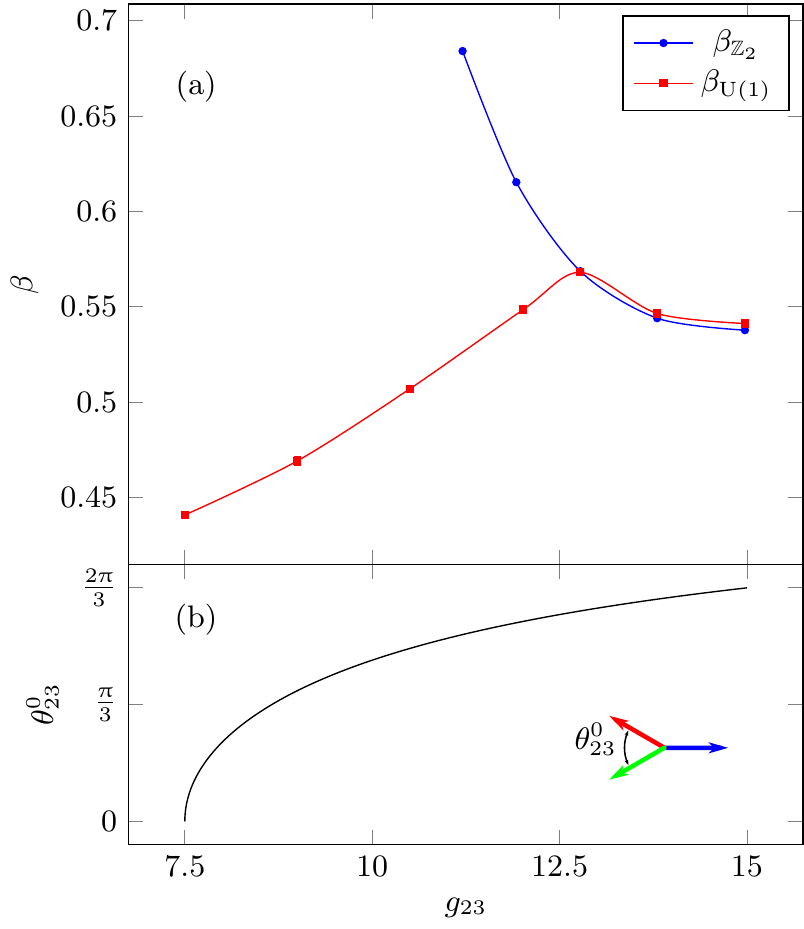}
\caption{(a) The phase diagram for the three-band model with unequal Josephson couplings. We set $g_{12} = g_{13} = 15$ and varied $g_{23}$. Error bars 
are smaller than symbol sizes. Lines are guides to the eye. (b) {The phase difference $\theta^0_{23}$ between band 2 and 3 in the ground state, as 
defined in the phase vector inset}. $\theta^0_{23} = 0$ for $g_{23} < 7.5$. $\theta^0_{12} = \theta^0_{13} = (2\cpi - \theta^0_{23})/2$ for all 
$g_{23}$.\label{fig:pd_gen_asym}}
\end{figure}

Next, we consider the case of a more general model where the Josephson couplings are different. By tuning some of the Josephson couplings one can make the difference 
between two out of three phases arbitrarily small in the BTRS ground state. This also implies that the energy  of $\groupZ{2}$ domain walls can be made arbitrarily 
small. Thus, one can interchange critical temperatures of $\groupU{1}$ and $\groupZ{2}$ phase transitions. Moreover, inclusion of fluctuations can in a certain limit
dramatically suppress the critical temperature of the $\groupZ{2}$ phase transition. Results from Monte-Carlo simulations shown in Fig. \ref{fig:pd_gen_asym} display 
such behavior.

Finally, consider the effect of a finite penetration length. As can be seen from Eq. \eqref{ps}, the gauge field couples only to the $\groupU{1}$ 
sector of the model, making the $\groupU{1}$ symmetry local. It also makes the energy of $(1,1,1)$ and $(-1,-1,-1) $ composite vortices finite \cite{cp22}.
As a result, at any finite temperature, there is a finite probability of exciting  such topological defects, which from a formal viewpoint suppresses 
superconductivity at finite temperature in the thermodynamic limit. In a real experiment on a finite system, with large but finite penetration length, 
this physics manifests itself as a conversion of a BKT transition to a broad crossover which takes place at lower characteristic temperatures than 
the  $\groupU{1}$ phase transition in  the global $\groupUZ$ model. On the other hand, since the $\groupZ{2}$ phase transition is not directly affected 
by this coupling, the $\groupZ{2}$ ordered non-superconducting state persists. Thus, in the thermodynamic limit a superconducting system with finite penetration 
length features $\groupUZ$ superconductivity at zero temperature, while at any nonzero temperature it resides in a $\groupZ{2}$ metallic state, 
up to the temperature where the  $\groupZ{2}$-symmetry is restored. In other words, a finite penetration length increases the phase space of a 
metallic state with broken time-reversal symmetry. The arguments above carry over to three dimensions as well. 

In conclusion, we have studied  the phase diagram of three band superconductors with spontaneously broken time reversal symmetry due to frustrated 
interband Josephson-couplings, beyond mean field approximation. {We have found that there is a new fluctuation-induced non-superconducting state 
which also exhibits a spontaneously broken time reversal symmetry, associated with persistent interband Josephson currents in $\v{k}$-space. This 
state is distinct from an ordinary metallic state where there is no such broken symmetry.}
Experimentally, it can be distinguished from superconducting and ordinary normal states by a combination of local (e.g. tunneling) and transport 
measurements. Another way of possibly detecting this state would be by observing an Onsager anomaly in the 
specific heat in the normal state.
These predictions could also be used to verify if \ce{Ba_{$1-x$}K_{$x$}Fe2As2} 
breaks time reversal symmetry at certain doping. A related phase should also exist in other superconductors which break time reversal 
symmetry  \cite{Weston2013}, as well as in interacting multicomponent Bose condensates.

\begin{acknowledgments}
We thank J. Carlstr\"om and E. V. Herland for fruitful discussions and feedback. T.A.B. thanks NTNU for financial support, and the Norwegian consortium 
for high-performance computing (NOTUR) for computer time and technical support. A.S. was supported by the Research Council of Norway, through Grants 
205591/V20 and 216700/F20. E.B. was supported by Knut and Alice Wallenberg Foundation through the Royal Swedish Academy of Sciences Fellowship, Swedish
Research Council and by the National Science Foundation CAREER Award No. DMR-0955902. EB and AS thank the Aspen Center for Physics for its hospitality 
during this work. 
\end{acknowledgments}

 \bibliography{references}
 
\end{document}